\documentclass[10pt,journal, twocolumn]{IEEEtran}

\usepackage{algorithm}
\usepackage{algpseudocode}
\usepackage{enumitem}
\usepackage{mathtools,amsfonts,amssymb,amsthm, bm, nccmath}
\usepackage{xcolor}
\usepackage{url}
\newcommand{\er}{Erd\H{o}s-R\'enyi }

\theoremstyle{definition}

\theoremstyle{remark}

\algtext*{EndWhile}% Remove "end while" text
\algtext*{EndIf}% Remove "end if" text
\algtext*{EndFor}% Remove "end For" text
\algtext*{EndFunction}% Remove "end Function" text

\begin{document}

\title{Entanglement Routing Based on Fidelity Curves}

\author{
    Bruno C. Coutinho, Raul Monteiro, Luís Bugalho, Francisco A. Monteiro, ~\IEEEmembership{Member, IEEE}
      
    \thanks{Bruno Coutinho is with Instituto de Telecomunicações, Lisbon, Portugal, e-mail: bruno.coutinho@lx.it.pt.}
    
    \thanks{Raul Monteiro is with Instituto de Telecomunicações, and Instituto Superior T\'{e}cnico, Universidade de Lisboa, Lisbon, Portugal.}
    
    \thanks{Luís Bugalho is with Physics of Information and Quantum Technologies Group, Centro de Física e Engenharia de Materiais Avançados (CeFEMA), and Instituto Superior T\'{e}cnico, Universidade de Lisboa, Lisbon, Portugal, and LIP6, Sorbonne Université, CNRS, Paris, France}
    
    \thanks{Francisco A. Monteiro is with Instituto de Telecomunicações, and ISCTE - Instituto Universitário de Lisboa, Lisbon, Portugal, e-mail: francisco.monteiro@lx.it.pt.}
}

\maketitle
    
\begin{abstract}
How to efficiently distribute entanglement over large-scale quantum networks is still an open problem that greatly depends on the technology considered. In this work, we consider quantum networks where each link is characterized by a trade-off between the entanglement generation rate and fidelity. For such networks, we look at the two following problems: the one of finding the best path to connect any two given nodes, and the problem of finding the best starting node in order to connect three nodes in the network multipartite entanglement routing. Two entanglement distribution models are considered: one where entangled qubits are distributed one at a time, and a flow model where a large number of entangled qubits are distributed simultaneously. The paper proposes of a quite general methodology that uses continuous fidelity curves (i.e., entanglement generation fidelity vs. rate) as the main routing metric. Combined with multi-objective path-finding algorithms, the fidelity curves describing each link allow finding a set of paths that maximize both the end-to-end fidelity and the entanglement generation rate. For the link models and networks considered, it is proven that the algorithm always converges to the optimal solution. It is also shown through simulation that the execution time grows polynomially with the number of network nodes (growing with a power between $1$ and $1.4$, depending on the network).
\end{abstract}


\section{Introduction}
The quantum Internet has the potential to allow capabilities and services that would be impossible in classical networks. To name a few, it opens doors to theoretically fully secure communications, enhanced sensing, and distributed quantum computation~\cite{van2014quantum}. A quantum Internet aims to create entanglement between remote nodes, a type of correlation between different parties with no classical analogue~\cite{nielsen00}. The typical model for a quantum internet is a network where the nodes store qubits in quantum memories, and the links between two nodes represent a quantum channel capable of creating quantum entanglement between the qubits stored at the nodes at both ends of the link. As often assumed, one assumes that it is possible to apply two-qubit gates between any two qubits inside the same node, and that each node has a fixed number of memories associated with each of the channels connected to it.

Different metrics can be used to characterize a quantum link. A popular and useful one is the fidelity attainable by a link; however, that fidelity can depend on other variables closely related to the technology used to sustain quantum entanglement. For example, one may apply link purification such that from a larger number of qubits having a low fidelity one can create a lower number of qubits holding a higher fidelity. The technology considered in this paper is based on photon entanglement, and, in this case, the achieved fidelity of a qubit pair is chiefly dependent on the number of photons (number of ``clicks'') generated, such that one can characterize the link by a trade-off curve between entanglement probability and the obtained fidelity in the entangled qubit pair.

\begin{figure}[tb]
    \centering
    \includegraphics[width=0.4\textwidth]{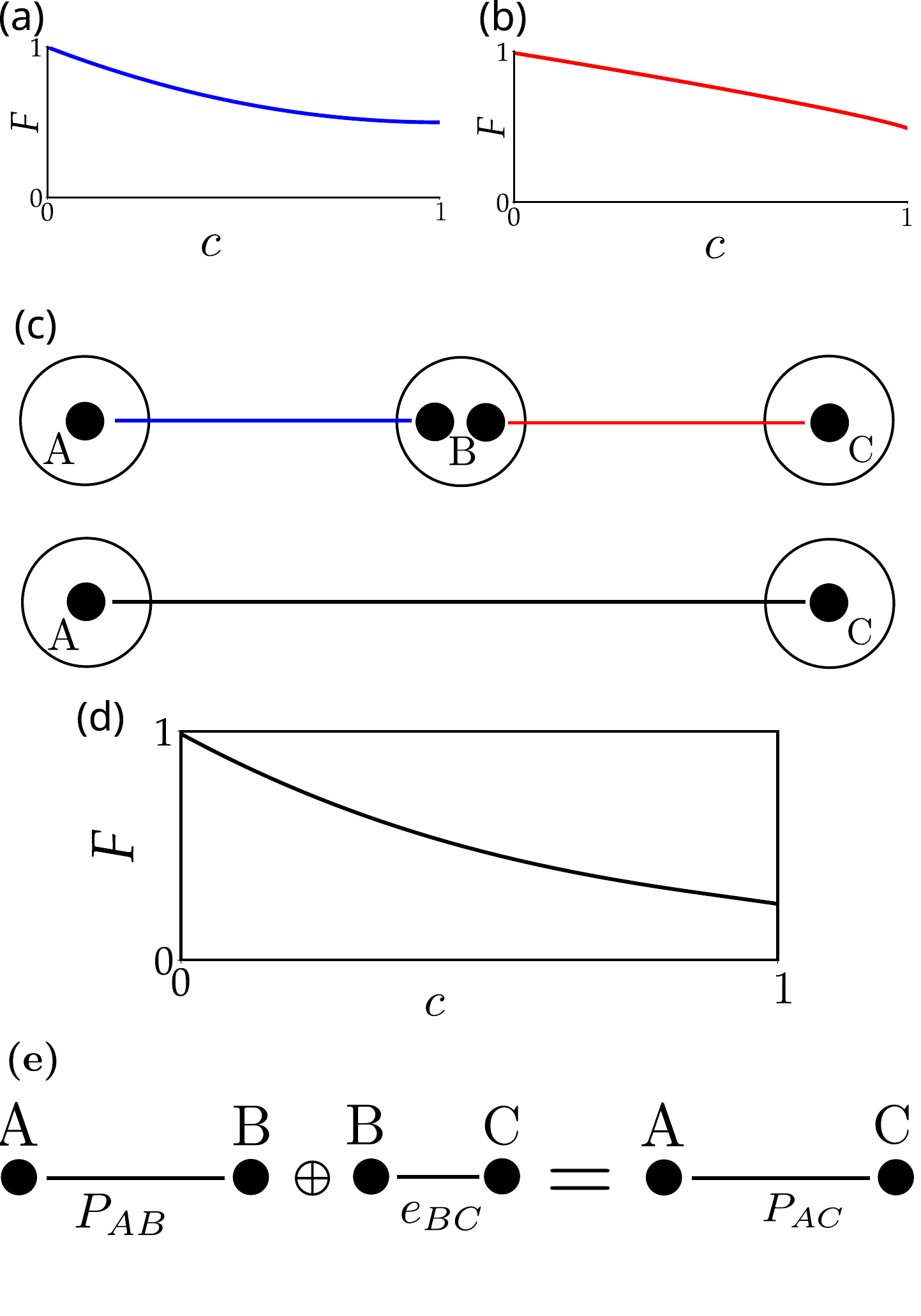}
    \caption{\textbf{Entanglement distribution.} (a-b) Shows fidelity as a function of the entanglement generation probability $p$ (and the equivalent in terms of capacity $c$, for $n_e=1$ ebit) for two links. Two fidelity curves are depicted: one for the link between nodes $A$ and $B$ (a) and the other for the link between nodes $B$ and $C$ (b). (c) shows a simpler quantum repeater setup. $A$, $B$, and $C$ represent three nodes: node $B$ contains two qubits, with one entangled with a qubit contained in node $A$ and the other in node $C$. Applying a swapping protocol at the two qubits contained in node $B$ creates an entanglement between the qubits in nodes $A$ and $C$ according to the fidelity curve shown in (d). The end-to-end link capacity will depend on the distribution model considered, which can be either the single ebit distribution model or the flow model. In (e) this process is described in a more formal way, where the concatenation of path $P_{AB}$ with a network link $e_{BC}$ creates a path between node $A$ and $C$ with the fidelity curve shown in (d).}
    \label{fig1}
\end{figure}

Remote entanglement generation involving distant nodes that are not directly connected can be achieved through the swapping mechanism~\cite{caleffi2017optimal, Pirandola2019, chakraborty2020entanglement, Bugalho2023, ghaderibaneh2022efficient, Pant2019} applied at several middle nodes. This procedure creates entangled qubit pairs (ebits) between nodes (as exemplified in Fig.~\ref{fig1}). By using several swapping operations it is possible to create entanglement between any two nodes in a network, provided that a path connecting the two nodes exists, and also that the end-to-end quality of the entangled pair generated meets the requirements of the specific application.

Finding the best route to distribute entanglement has proven to be a non-trivial problem \cite{caleffi2017optimal, Pirandola2019, chakraborty2020entanglement, Bugalho2023, ghaderibaneh2022efficient, Pant2019}. Caleffi et al. \cite{caleffi2017optimal} studied a single-qubit entanglement generation model for bipartite entanglement (i.e., between only two nodes). In their model, the entangled qubits start as maximally entangled and subsequently lose coherence over time due to imperfect quantum memories. They showed that it is not possible to use Dijkstra's algorithm to find the route that maximizes the entanglement distribution rate, and proposed an algorithm to exactly solve this problem. Although not directly stated, the execution time of the algorithm in \cite{caleffi2017optimal} grows supra-exponentially with the number of nodes in the network, however, the authors point out that for small quantum networks that does not constitute a problem.

Pirandola et al.\cite{Pirandola2019} looked at bipartite entanglement networks based on the theoretical upper bounds for the channel capacity. In a regime where entanglement distribution is close to its theoretical upper bound, that work showed that in such a regime Dijkstra's algorithm can be used to find the path that maximizes the entanglement distribution rate between nodes, and the max-flow min-cut theorem can be used to find the maximum rate between two nodes using multiple-path entanglement distribution. Both problems can be solved in polynomial time, and this approach was later generalized to include the multipartite entanglement distribution of GHZ-states\cite{Bauml2020}.

Chakraborty et al.~\cite{chakraborty2020entanglement} studied bipartite entanglement distribution in a flow model in which the ebits all have the same fidelity (quality of the entanglement), but each link has different capacities. Here, we note that a \textit{flow model} is one in which multiple ebits are attempted to be established simultaneously. Those authors proposed a multicommodity flow algorithm that can find the optimal flow for bipartite entanglement distribution between two sets of nodes in polynomial time.

Bugalho et al. considered in \cite{Bugalho2023} bipartite and multipartite entanglement distribution assuming a single-qubit generation model and considering that not all links have the same fidelity or entanglement generation probabilities, and combined that with imperfect quantum memories. That work considered a single source model (i.e., only one source node trying to establish entanglement with all the other nodes) and proposed a multi-objective routing algorithm to find the routes that simultaneously maximize the bipartite or multipartite entanglement generation rate and the fidelity. The algorithm is NP-hard, even though the authors showed that, for the networks analyzed, the algorithm always converges in polynomial time.

Ghaderibaneh et al.~\cite{ghaderibaneh2022efficient} focused on a bipartite entanglement distribution model with a non-deterministic swap. As a consequence of that, the order in which each of the swapping operations is applied impacts the rate of entanglement distribution. They proposed a polynomial-time algorithm to find both the optimal path and the optimal swapping order.

Finally, Santos et al.~\cite{sarasantos2023} proposed a multi-objective algorithm for bipartite entanglement distribution between two nodes, considering a network based on an asymptotic description of quantum repeaters and purification protocols, however, with no detail given on the specific protocol.

All of these works represent steps forward not only regarding routing techniques in the quantum networks but also in how detailed the network models are. That being said, we are still far from having a network model that fully incorporates the intricacies of entanglement generation and distribution.

\section{Entanglement Generation}

This paper advances the research on quantum networks by introducing an highly versatile routing approach based on fidelity curves, which can be utilized in conjunction with purification protocols including capacity-achieving purification ones~\cite{Roque2023}. The fidelity of an ebit can be quantified as the distance between the quantum state of the ebit in question and the state of a maximally entangled ebit~\cite{nielsen00}. Typically, this fidelity can be manipulated by altering the entanglement distribution rate.
Two examples illustrate this trade-off. First, consider the setup proposed by \cite{Childress:2005}, where entanglement between qubits is generated using laser pulses. Increasing the duration of these pulses raises the probability of generating an ebit but reduces its quality, as shown in Fig. 1(a,b). Consequently, if a link's objective is to generate a large number of qubits with less emphasis on quality, longer laser pulses would be used. Conversely, if the link requires fewer ebits but with high fidelity, shorter pulses would be preferred.
A second example involves the use of quantum error correction \cite{Cruz_Access_2023} and/or quantum purification to enhance ebit fidelity at the cost of reduced entanglement generation rate \cite{Roque2023}. For example, if one generates $n$ ebits in a link with fidelity $f_{\rm in}$, one can employ quantum purification or error correction to transform these into $k<n$ ebits with a higher fidelity $f_{\rm t}>f_{\rm in}$. Our approach can incorporate any purification protocols that occur between two adjacent nodes, but a particularly apt illustration of the trade-offs involved in such protocols is hash-based purification \cite{Roque2023}. This method offers two key advantages: first, it enables the generation of a code that, for a given number of input ebits $n$, can output any number of $k$ ebits. Second, as the number of input qubits approaches infinity, the fidelity of the output ebits tends towards the Hamming bound – the maximum fidelity possible for degenerate codes under these conditions~\cite{Roque2023}.
While hash-based methods for purification and error correction have traditionally been computationally intensive, recent advances have mitigated this challenge. New methods based on guessing random additive noise decoding (GRAND) \cite{Duffy_Medard_TIT_2019}, such as quantum-GRAND \cite{Cruz_Access_2023}, and purification-GRAND \cite{Roque2023}, now allow for the error correction and/or purification of up to 200 ebits.
Ultimately, whether changing laser pulses duration during the entanglement generation process, quantum error correction and purification post processing, or a combination of both, we obtain a fidelity vs. rate curve. This curve represents the fidelity of the generated entangled pairs for a given rate and it is a general tool that can fit a larger of link models. In some sense, it is analogous to the block error rate (BLER) curves used in classical communications.

\begin{figure}[tb]
    \centering
    \includegraphics[width=0.45\textwidth]{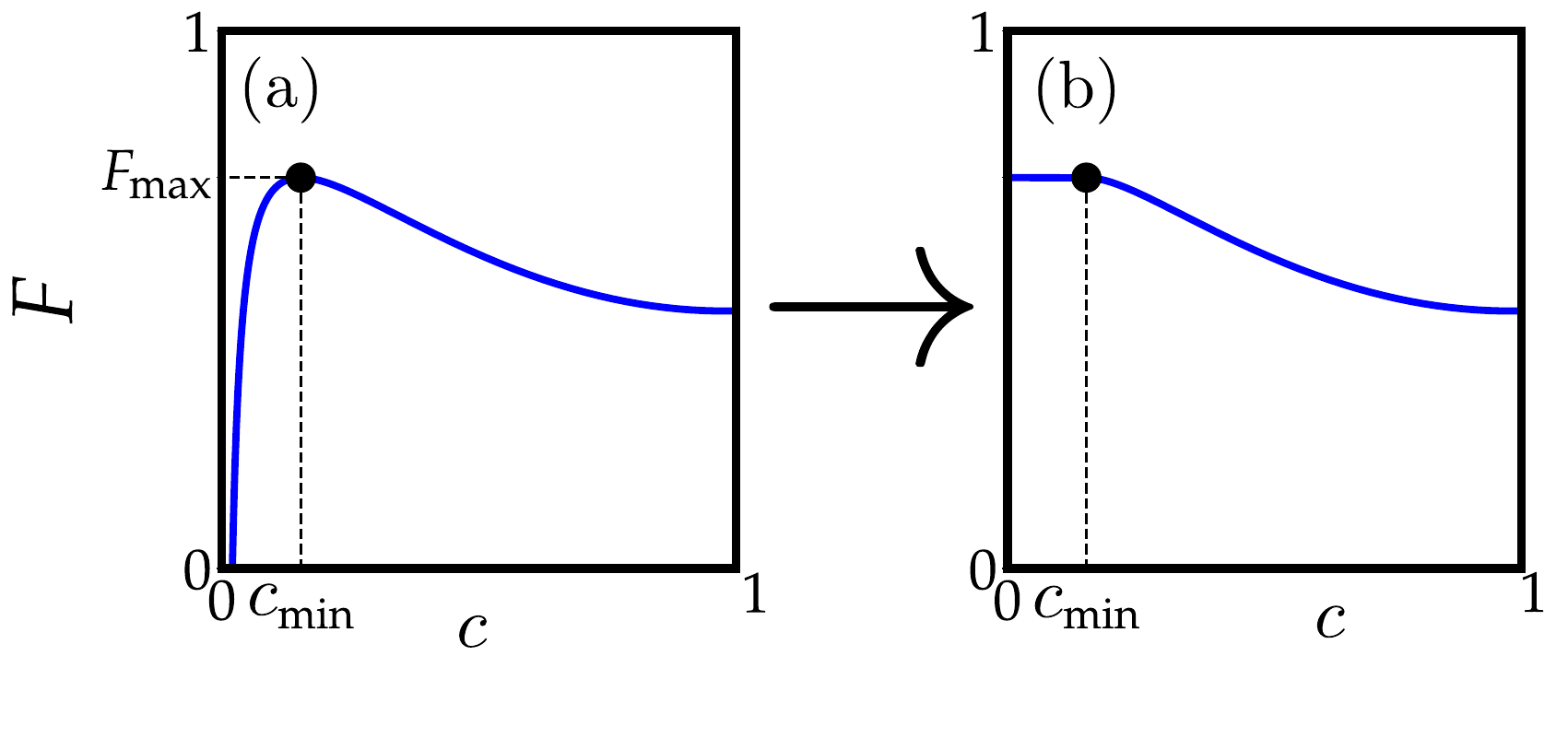}
    \caption{\textbf{Non-monotonically decreasing fidelity curves.} Figure (a) shows a non-monotonically decreasing fidelity due to the existence of dark counts in the link model proposed by \cite{Childress:2005}. In this example, for a low entanglement generation rate ($c<c_{\rm min}$), the dark counts dominate, causing a fidelity reduction. This problem can be fixed if for each value of $c_{ij}$, we select the largest fidelity with an entanglement rate above $c_{ij}$, and thus one obtains a monotonically decreasing function, shown in (b).}
    \label{fig2}
\end{figure}

The are situations where the fidelity curves are not monotonically decreasing functions, such as when dark counts are present \cite{Childress:2005}. A dark count occurs when the detector registers the arrival of a photon despite none had been emitted. If the emission probability is low, those events can become dominant and reduce the quality of the generated ebit. Fig.~\ref{fig2} depicts such phenomena. For any value $c_{e} < c_{\rm min}$, our toy example can consider an artificial monotonically decreasing function where, for $c_{e} < c_{\rm min}$ by defining $F_{e}(c_{e}) \rightarrow F_{e}(c_{\rm min})$. If multiple monotonically decreasing regions exist, the fidelity at a given capacity can always be replaced by the highest fidelity achievable at any higher capacity.
In practice, for each value of $c$ one selects the largest fidelity with an entanglement rate above or equal to $c$.

\section{Entanglement propagation}

The previous section only concerned entanglement generation between nodes that are directly connected. As mentioned before, it is also possible to generate entanglement between nodes that are not directly connected, provided that there is a path connecting them. Let us consider the simple example in Fig.~\ref{fig1}(c): although node $A$ and $C$ do not share an ebit, it is possible to generate one using a swap operation between the two qubits in node $B$~\cite{caleffi2017optimal, Pirandola2019, chakraborty2020entanglement, Bugalho2023, ghaderibaneh2022efficient, Pant2019}. 
This operation comes with a cost in terms of the quality of the entanglement generated, as seen in Fig.~\ref{fig1}(d). The fidelity curve for the entanglement generated between $A$ and $C$ (Fig.~\ref{fig1}(d)) is considerably worse than the fidelity curves for the entanglement generated between $A$ and $B$, and $B$ and $C$ (Fig.~\ref{fig1}(a-b)). How to compute such fidelity curves is highly dependent on the type of noise affecting the qubits, and also on the exact procedure used to create a connection between nodes $A$ and $B$.
It is often assumed that the noise affecting ebits can be described by a uniform depolarizing channel originating in Werner states~\cite{Bugalho2023}, whose quality is often characterized by the Werner parameter $\gamma$, related to the fidelity as $\gamma = (4F-1)/3$. Werner states have the convenient property that swapping two or more Werners generates another Werner state with a fidelity that is rather easy to calculate: the end-to-end $\gamma$ parameter is obtained by the product of the $\gamma$ parameters of each of its links \cite{Bugalho2023}.   
For the example, let us consider Fig.~(\ref{fig1}), where the entangled states between the nodes $A$ and $B$, and $B$ and $C$, the $\gamma$  parameter of the Werner state connecting path between A and C is the just the multiplication of the $\gamma$ paremter of the Werner states connecting and $A$ and $B$ and $B$ and $C$~\cite{Bugalho2023}, and with it we can then easy find the fidelity. 
In the rest of the paper, the term ``fidelity curves'' will be used to denote both fidelity and Werner parameter vs. channel capacity, given they linear relation.
%In the following subsections, we look at the entanglement distribution rate, where we consider two distribution models: a single ebit distribution model \cite{sarasantos2023,Bugalho2023}, and a flow model~\cite{chakraborty2020entanglement,dai2020optimal}.

\subsection{Single ebit distribution model}

In the single ebit distribution model all links in a path $P$ try to generate one ebit simultaneously, and the entanglement is only generated between the source and target nodes if all links successfully generate entanglement. This method is especially useful for situations where the quality of the quantum memories is low; therefore, the entanglement between the ebits rapidly decays with time. Because all entanglement generation is done simultaneously, the links only need to store the entanglement in a quantum memory for the time necessary to apply the swapping protocol. Let us then consider a path $P$ in a network where each link $e$ generates one ebit per fixed time interval $t$ with a probability $p_e$, corresponding to an average capacity $c_e=p_e$. Note that because only one qubit is generated per time slot, the maximum rate (i.e., the capacity) corresponds to the probability $p_e$.
%The end-to-end Werner parameter and the end-to-end rate between a source node and a target node are respectively given by
Different combinations of link capacities, $\{c_e\}$, can result in the same end-to-end capacity $c^{\rm sg}$, and therefore the end-to-end Werner parameter is not a function of the end-to-end capacity. This problem can be overcome by always considering the combination of link capacities that maximizes the Werner parameter (and consequently the fidelity) for a given end-to-end capacity.  or a path $P$ defined as a set of links $e$, the Werner parameter that maximizes the end-to-end fidelity can be written as:

\begin{align}
    &\gamma^{\rm sg}(c^{\rm sg})=\max_{c_e}\prod_{e \in P} \gamma_{e}(c_e)\\
    &c^{\rm sg}=\prod_{e \in P} c_e
    \label{eq:final_gamma},
\end{align} 
where $c_e$ are the parameters associated with each link $e$ in the path $P$, and  $\gamma_{e}(c_e)$ represents the Werner parameter for each link. This optimization can be performed sequentially as shown in Appendix \ref{optimality_0}.

The limitation of this model is that both the fidelity and the generation probability decay with the length of the path. For this reason, the single distribution protocol might not be the preferred one for entanglement generation over long distances.

\subsection{Flow distribution model}

The previous limitation can be ameliorated in the so-called flow distribution model, in which multiple ebits are generated simultaneously at each node. A link, $e$, contains $N_e$ ebits, the capacity of each link is defined as $C_e=N_e p_e$, and the capacity of a path is given by the smallest capacity of any of the links on the path. When generating a large number of qubits simultaneously, the statistical fluctuations of the entanglement generation process are so low that can be ignored. For practical purposes, one can consider that a fixed number of ebits, $C_e$, is generated at each entanglement attempt. All swapping operations are executed in parallel, and therefore the method is useful when the quality of the quantum memories is low and a large number of ebits can be generated simultaneously. Likewise the first model, because all entanglements are carried out simultaneously, the links only need to store the entanglement in a quantum memory for the time needed to apply the swapping operation.
%Once again, the decay in fidelity that takes place during such time will just affect the parameter $\beta$ in the fidelity curves. Although not directly stated in the model, the entanglement distribution rate still decays with distance.
%Given that the fidelity and the capacity of each link are correlated (see Fig.~\ref{fig1}(a,b))...
Since the ebits' end-to-end fidelity decays with distance, one has to reduce the entanglement distribution rate as the distance increases. In general, the fidelity of each ebit in a path decays with its capacity, and the end-to-end rate of entanglement generation is bottlenecked by the smallest capacity among its links. If the fidelity is a monotonically decreasing function with respect to the links' capacity, the optimal solution is for all links to operate with a capacity equal to the end-to-end entanglement distribution rate, and use the fact that the end-to-end Werner parameter is just the multiplication of the $\gamma$ parameter of all the links in the path, i.e., for a path $P$ defined as a set of links $e$, its Werner parameter is given by (as proven in Appendix \ref{optimality_1})
\begin{align}
    \gamma^{\rm flow}_{P}(c) &= \prod_{e \in P} \gamma_{e}(c), \label{eq:final_gamma}
\end{align}
where $\gamma_{e}(c)$ is the Werner parameter of a link $e$ for a certain capacity $c$.

\section{Routing metrics, Isotonicity and Monotonicity}

Formal algebras are useful tools to understand routing, especially when convergence to the optimal solution is of interest~\cite{Bugalho2023}. Although an extensive explanation of routing algebras is outside the scope of this work, we will introduce some useful concepts. The propagation of entanglement described in the previous section is often called path extension and is denoted by $\bigoplus$. For example, in Fig.\ref{fig1}(c), the path between $A$ and $B$, $P_{AB}$, is expanded to connect $A$ and $C$ using the link $e_{BC}$. In routing theory, this is termed as a concatenation between $P_{AB}$ and the link $e_{BC}$, represented as $P_{AC}=P_{AB}\bigoplus e_{BC}$, as depicted in Fig.\ref{fig1}(e). Routing metrics are another important concept in routing theory; they consist of a set of parameters that characterize each path and allow one to compare two paths connecting the same nodes, deciding if one is preferable to the other, or if neither is necessarily better. A routing metric can be a real number (i.e., a simple weight), or a multidimensional quantity, for example, a set of distances or objectives. As stated earlier, the main contribution of this work is the proposal that fidelity curves can be good routing metrics and can be used to find the optimal path between two nodes in a network.

\begin{figure}[tb]
    \centering
    \includegraphics[width=0.4\textwidth]{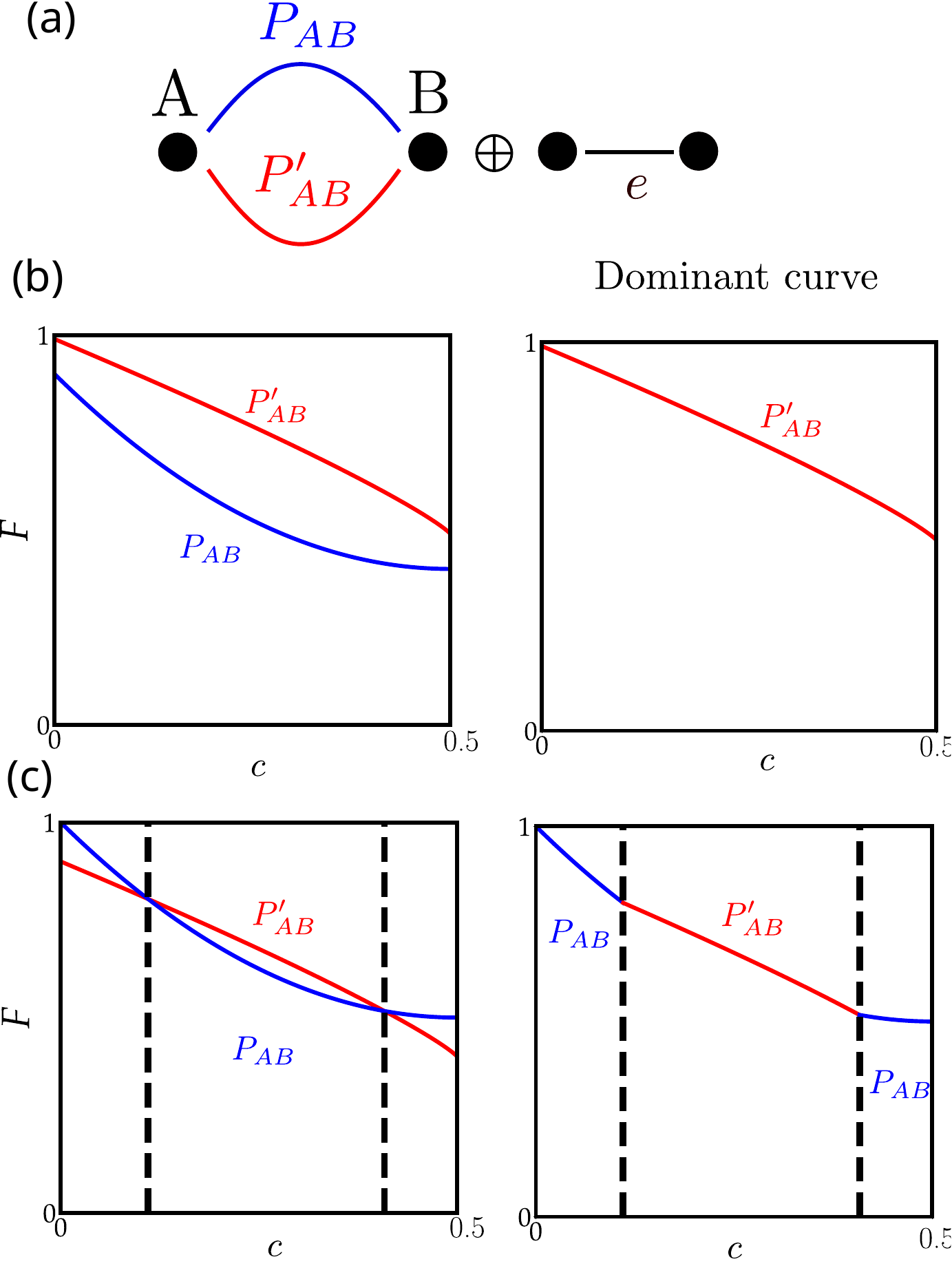}
    \caption{\textbf{Comparison between paths and dominance.} Let one suppose that there are two paths, $P_{AB}$ and $P'_{AB}$, between node $A$ and $B$, as represented in (a). $P'_{AB}$ dominates $P_{AB}$ if, independently of the properties of the network segment $e$, $P'_{AB} \bigoplus e$ dominates $P_{AB} \bigoplus e$. (b)-left shows an example where $P'_{AB}$ dominates $P_{AB}$, therefore the dominant curve, (b)-left, is composed of $P'_{AB}$ only. Figure (b)-right shows an example where neither path $P'_{AB}$ nor $P_{AB}$ dominates.} \label{fig3}
\end{figure}

It is known that if a routing metric is both isotonic and monotonic, a multi-objective optimization algorithm will always converge to the set of optimal solutions~\cite{Bugalho2023, sarasantos2023}. \textit{Monotonicity} means that when a path is extended, our metric will either always increase or always decrease. To introduce \textit{isotonicity}, consider two paths, $P$ and $P'$, connecting the same end-nodes, and a point on the fidelity curve for both paths such that $\gamma_P(c) \leq \gamma_{P'}(c)$. When extending both paths with an extra edge, as exemplified in Fig.\ref{fig3}(a), if for $P\oplus e$ and $P'\oplus e$ we still have $\gamma_{P\oplus e}(c) \leq \gamma_{P' \oplus e}(c)$, then the metric is isotonic for the capacity $c$. Introducing the concept of dominance allows us to consider the entire fidelity curve. In our work, we consider that one path dominates another (denoted as $P' \textbf{ D } P$) if the fidelity curves associated with paths $P'$ and $P$ are distinct and the fidelity curve of $P'$ is above or equal to the fidelity curve of $P$ for any capacity $c$ in the relevant domain, as exemplified in Fig.~\ref{fig3}(b). If, for every value of the capacity, the metrics are monotonic and isotonic, then the dominance relation is inherited when extending the paths, and the algorithm is able to find the optimal solution. Since the algorithm merges curves from different paths into one, keeping only the highest value from each (as shown in Fig.~\ref{fig3}(c)), then the isotonicity of the curve and dominance inheritance for every point of the curve have the same meaning. Fidelity curves are trivially isotonic and monotonic. Because the end-to-end fidelity curves in both the single ebit and flow distribution models are composed of multiplications and max/min functions, it is straightforward to prove that these curves are both isotonic and monotonic. For more complex distribution models, special attention should be paid to demonstrate whether the fidelity curves are isotonic and monotonic. This will need to be evaluated on a case-by-case basis, and modifications of this approach might be required.

\subsection{Isotonicity and monotonicity for the single ebit distribution model}

Let us divide the proof that the fidelity curves in the parallel entanglement distribution model are monotonic and isotonic metrics.
First, it is important to note that one always obtains the same fidelity curve independently of the order of the concatenations, as shown in Appendix~\ref{optimality_0}. Given this, in order to prove monotonicity, we need to show that the fidelity curve $\gamma_{P}(c)$ of a path $P$ is above or equal to the fidelity curve, $\gamma_{P'}(c)$, of a path $P' := P \oplus e$, where $e$ is a single link with a fidelity curve $\gamma_{e}(c)$. We can then write (see Appendix~\ref{optimality_0} for details),
\begin{align}
\gamma_{P'}(c) = \max_{c_e' \in P'} \left[\prod_{e' \in P}\gamma_{P}(c_e')\right]=\max_{c_P,c_e} [\gamma_{P}(c_P) \gamma_{e}(c_e)],
\end{align}
with the constraint that the capacity of path $P'$ is $c$. Therefore, the product between the capacity of path $P$, $c_P$, and the capacity of the extended link, $c_e$, needs to be equal to $c$ (that is, $c_P c_e = c$). By assuming that fidelity curves are monotonically decreasing functions, one can write (see Appendix~\ref{optimality_0} for details)
\begin{align}
\gamma_{P'}(c) = \max_{c_e} [\gamma_{P}(c/c_e) \gamma_{e}(c_e)]\leq \gamma_{P}(c),
\end{align}
where we use the fact that $\gamma_{e}(c_e)$ is at most equal to one, and therefore $c_P = c/c_e \geq c$, proving that the fidelity curves in the single ebit entanglement distribution model are monotonically decreasing metrics.

We now move on to proving isotonicity. Let us consider two paths $P_1$ and $P_2$, and the respective extensions, $P'_1 = P_1 \oplus e$, and $P'_2 = P_2 \oplus e$, with fidelity curves,
\begin{align}
\gamma_{P_1'}(c) =& \max_{c_{P},c_e} [\gamma_{P_1}(c_P) \gamma_{e}(c_e)]\\
\gamma_{P_2'}(c) =& \max_{c_{P},c_e} [\gamma_{P_2}(c_P) \gamma_{e}(c_e)],
\end{align}
with $c = c_P c_e$.
We need to prove that if $P_1 \textbf{ D } P_2$, it implies that $P_1' \textbf{ D } P_2'$, i.e., if the fidelity curve of path $P_1$ is equal to or above the fidelity curve of path $P_2$ for all capacities, then the fidelity curve of path $P'_1$ will also be equal to or above the fidelity curve of path $P'_2$ for all capacities. One can prove this by contradiction. Let us suppose that there is a capacity value $c$ such that the fidelity of path $P'_2$ is larger than the fidelity of path $P'_1$, i.e.,
\begin{align}
\max_{c_e^*} [\gamma_{P_2}(c/c_e^*) \gamma_{e}(c_e^*)] >
\max_{c_e^{**}} [\gamma_{P_1}(c/c_e^{**}) \gamma_{e}(c_e^{**})].
\end{align}
This equation establishes that the optimal fidelity for capacity $c$ in path $P_2$ is larger than the optimal fidelity for capacity $c$ in path $P_1$. In turn, this also implies that for the value of $c_e^*$ which maximizes the fidelity in path $P_2$, we get that:
\begin{align}
\gamma_{P_2}(c/c_e^*) \gamma_{e}(c_e^*) &> \gamma_{P_1}(c/c_e^*) \gamma_{e}(c_e^*) \\
\implies \gamma_{P_2}(c/c_e^*) &> \gamma_{P_1}(c/c_e^*),
\end{align}
\noindent which is a contradiction, since we assumed a priori that $P_1$ dominates $P_2$. We can therefore conclude that fidelity curves in the parallel entanglement distribution model are isotonic.

\subsection{Isotonicity and monotonicity for the flow entanglement distribution model}

The proof that the fidelity curves are isotonic and monotonic in the flow model is more straightforward. Let us consider Eq.~(\ref{eq:final_gamma}) applied to a path $P$ that goes from node $i$ to node $j$ using one or more hops in between, and which is extended using a link between nodes $j$ and $k$, $e_{jk}$. In a more formal way, $P' = P \bigoplus e_{jk}$. For simplicity, we will drop the notation of the edges in link $e$. One obtains the fidelity curves,
\begin{align}
    \gamma_{P'}(c)=&\prod_{e' \in P'} \gamma_{e'}(c)= \\& \prod{e' \in P} \gamma_{e'}(c) \times \gamma_{e}
    =\gamma_{P}(c) \gamma_{e}(c).
\end{align} 
Given the fact $\gamma_{e}(c) \leq 1$, we can conclude that
\begin{align}
    &\gamma_{P'}(c) \leq \gamma_{P}(c),
\end{align} 
and from that, we conclude that the fidelity curves are monotonically decreasing metrics in relation to path extension.
Using the concept of dominance, one can show that in fact, these fidelity curves are also isotonic. Let us consider two paths, $P$ and $P'$, between the nodes $i$ and $j$, with the two paths characterized by $\gamma(c_{ij})$ and $\gamma'(c_{ij})$, respectively. Similarly, $P \oplus e$ and $P' \oplus e$ are given by
\begin{align}
    &\gamma_{P}(c) = \gamma(c) \gamma_{jk}(c) \\
   & \gamma_{P'}(c) = \gamma'(c) \gamma_{jk}(c) 
\end{align}
and
\begin{align}
    &\gamma'_{ik}(c)/\gamma_{ik}(c) = \gamma'_{ij}(c)/\gamma_{ij}(c).
\end{align}

If $P'$ dominates $P$ (i.e., $\gamma_{ij}(c)/\gamma_{ij}'(c) \geq 1$ for all values of $c$), then it is easy to see that $P' \oplus e$ also dominates $P \oplus e$ (i.e., $\gamma_{ik}(c)/\gamma_{ik}'(c) \geq 1$ for all values of $c$). Fig.~\ref{fig3}(c) shows a situation where neither $P$ nor $P'$ dominates. In such cases, the dominant curve is composed of parts of the curves of both paths.

\section{Multi-objective Routing Algorithm}
%In the last section, we studied the properties of fidelity curves as routing metrics and the importance of them being both isotonic and monotonic.
When both isotonicity and monotonicity hold, it is possible to devise a simple multi-objective routing algorithm \cite{Bugalho2023} that finds the best routes between a source node and all other nodes. First, one starts at a source node and computes the fidelity curves and a priority value for the links that connect to its neighboring nodes. All discovered nodes are added to a priority queue that orders them according to the priority value.
The algorithm then moves to the path with the highest priority that connects the source node to a node $i$. This path is then extended to connect the source node to all neighbors of node $i$, calculating both the fidelity curves and the priority values of these new paths.
Inevitably, at a certain point, the algorithm will find a node that already has an associated fidelity curve. In this situation, the algorithm has found an alternative path to reach the neighboring node and has to check whether this new path is dominated or not by the old one.
To do this, temporary registry is required to keep track of the optimal fidelity curve found for each node, alongside the respective paths (as represented in Fig.\ref{fig3}). If the new path is dominated by the old path (as exemplified in Fig.\ref{fig3}(b)), the algorithm stops and moves to the next node in the priority queue. Instead, if the new path is not dominated by the old one, the fidelity curve and the respective path are recomputed (see Fig.~\ref{fig3}(c)), and consequently, all paths that go through this neighboring node need to be re-evaluated. This is achieved by adding this neighboring node to the priority queue.
This process continues until the priority queue is empty. Note that we did not specify how the priority value of each node is computed simply because we do not know what the optimal way to compute such a value is. Prioritizing paths with a low likelihood of being dominated will reduce the number of recomputations and paths added to the priority queue. 

Although multi-objective routing algorithms often converge to the optimal solution in polynomial time, multi-objective routing problems are generally NP-hard \cite{bokler2017multiobjective}. Therefore, choosing a good priority value is crucial. In our case, we used the length of the path as our priority value. While the shortest paths are not necessarily the dominant paths, they serve as a good initial guess. This is because fidelity rapidly decays with path length, making it increasingly unlikely that longer paths will not be dominated by much shorter ones. The pseudo-code for a possible algorithm is shown in Algorithm~\ref{alg:SPT}.
%The {\color{blue} runtime of the protocol is evaluated numerically in the section \s}. %We leave a rigorous analytical study of the convergence of our approach for future work.

    \begin{algorithm}[tb]
    \caption{Bipartite routing: Source-to-all}\label{alg:SPT}
    \begin{algorithmic}[0]
         \State {\bf{Data structures and objects}}
        \State $\mathcal{A}:=$ a set containing all paths that are still to be visited, defined as  $P_i$ = $\left\{d(i), F_i,n_i\right\}$, ordered as a priority queue. The order is defined by an increasing $d(k)$, with the smallest distance at the top of the priority queue.
        \State 
        \State $\mathcal{F}^{\rm reg}:=$ is a set containing, for each node $n_i$, a function $\mathcal{F}^{\rm reg}_{n_i}$  representing the concatenation of all non-dominated paths from a source to node $n_i$, and the associated paths (as represented in Fig.~\ref{fig3}).
        \State 
        \Function{PathSelection}{$source$}
        \State  Initialize $\mathcal{A}$ as $\left\{0, F_0 , source\right\}$.
        \State  Initialize $\mathcal{F}$ as set of $F_{\rm null}$ and null paths.
        \While{$\mathcal{A}$ is not empty}
            \State Select path $P_i$ at the top of the priority queue
            \State Remove $P_i$  from $\mathcal{A}$.
            \If{$\mathcal{F}^{\rm reg}_{n_i} \textbf{ D } F_i  $}
            \State Do nothing
            \Else
                \State  Update $\mathcal{F}^{\rm reg}_{n_i}$ with a concatenation of $\mathcal{F}^{\rm reg}_{n_i}$   and \State $F_i$, as described in Fig.~\ref{fig3}.
                \For{each node $n_k$ (neighbour of node $n_i$)}
                    \State Add path $P_k$ to $\mathcal{A}$.
                    \EndFor
            \EndIf
        \EndWhile
        \EndFunction
        \State
        \Return $\mathcal{F}^{\rm reg}$
\State
\State{\bf{Abbreviations}}
\State $P_k$ is the extension of path $P_i$ to all nodes $n_k$, neighbours of node $n_i$. ${P_k :=P_i \oplus e_{ik}}$.
\State{$F_0$ is a function defined as $F_0=1$ for any capacity. $F_0:= 1~\forall~c$.}
\State{$F_{\rm null}$ is a function defined as $F_{\rm null}=0$ for any capacity.}
    \end{algorithmic}
\end{algorithm}
\section{Mutipartite entanglement distribution}

The proposal outlined so far can be integrated with the work developed in~\cite{Bugalho2023}, which focused on establishing simultaneous entanglement among more than two nodes. Specifically, we will target the distribution of three-partite entanglement, for which the scheme proposed in~\cite{Bugalho2023} is exact. Because the bipartite metrics considered in the present work are both monotonic and isotonic, one can ensure the optimality of the paths. This satisfies the first requirement for proving that the algorithm in~\cite{Bugalho2023} converges to the optimal solution.

The second step is to demonstrate that the multipartite metrics (specifically, the fidelity and rate metrics for the multipartite state) are monotonic and label-isotonic with respect to the paths. For these metrics, instead of adding links to extend paths, one should combine paths to form trees. In the particular case of a three-partite state, the tree connecting three terminals is also a star graph.
%The fidelity and rate metrics for a multipartite entangled state among three terminal nodes, $\mathcal{T} = {T_1, T_2, T_3}$, where $s$ is the source node, follow this structure.
Following the analysis in~\cite{Bugalho2023}, the fidelity of each star is both monotonic and label-isotonic. This means that if the fidelity of any individual path increases, the fidelity of the overall multipartite state also increases. Therefore, given that the $\gamma$ functions are monotonically decreasing with the rate, the most efficient way to generate the multipartite state is to maintain the same rate for each path. This approach maximizes the overall fidelity and minimizes losses due to differing rates. Consequently, for each capacity value, the metric is trivially monotonic and label-isotonic, enabling us to apply the previous dominance relation to the case of trees. Thus, our approach can be combined with the work of~\cite{Bugalho2023} to facilitate the distribution of not only bipartite entanglement but also multipartite entanglement.

Assuming that we already found the optimal fidelity curves between any two nodes in the network, one can easily find the fidelity curve associated with multipartite entanglement distribution for a given source node and any tree nodes~\cite{Bugalho2023}. One has to test all possible source nodes, which suggests that the run time of such an approach will grow linearly with the number of nodes in the network. Fig.~\ref{fig5} shows the run time as a function of the number of nodes in our implementation. In most cases, the run time is either linear or slightly sub-linear, with the exception of the single ebit distribution model in random geometric graphs, which exhibits an almost constant run time with respect to the number of nodes. This is due to the small number of valid paths connecting two nodes in such a model.

\section{Simulation results}

\begin{figure*}[tb]
    \centering
    \includegraphics[width=1\textwidth]{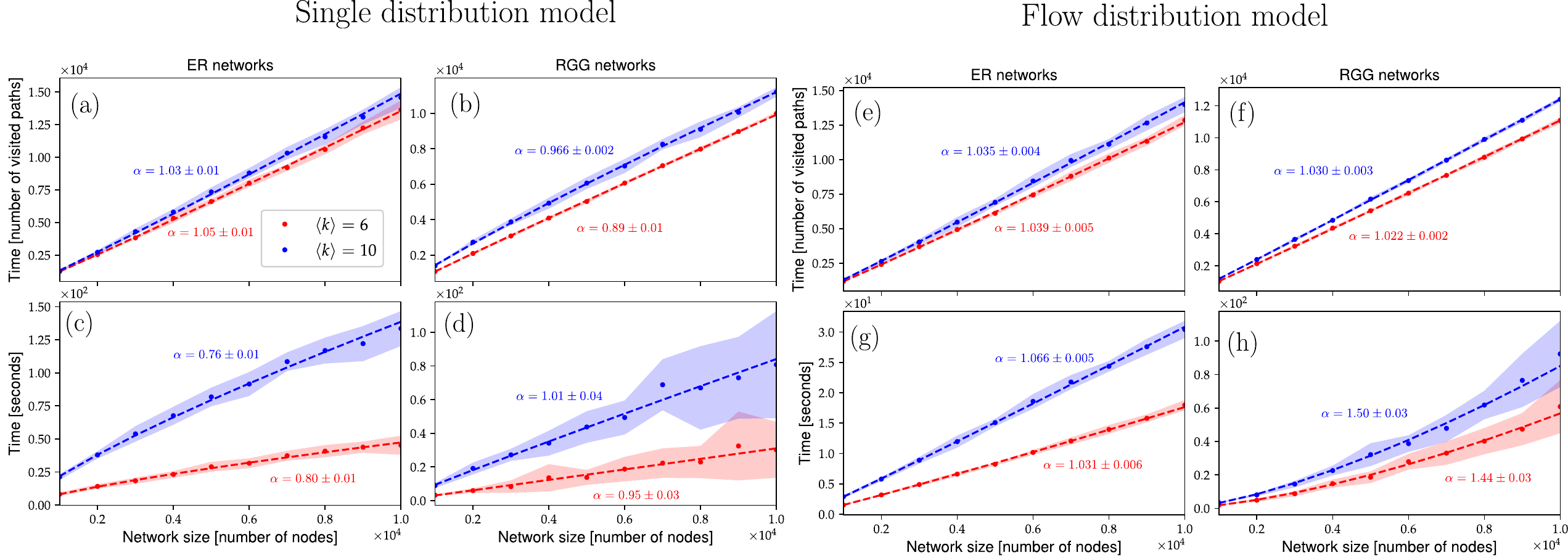}
    \caption{{\bf Bipartite routing runtime analysis}. Simulation of bipartite entanglement distribution for photonic quantum networks with an \er network (ER) and a random geometric graph (RGG) topology, an average degree $\langle k \rangle=6$ and $10$ (blue lines), for the single, (a-d), and flow, (e-h), distribution models. Each link follows the link model from \cite{Childress:2005} and his characterized by three parameters: $\epsilon$, the dark count $p_{\rm dark}$, and the minimum infidelity $\bar{F}_{max}$, randomly assigned from uniform distributions $U(0.3, 0.4)$, $U(0,{10}^{-3})$, and $U(0,{10}^{-3})$, respectively. (a-b) show the total number of visited paths as a function of the number of nodes in the network for an ER network and a RGG topology in the single distribution model, and (c-d) show the total execution time of the algorithm as a function of the number of nodes for the same  networks. (e-h) show the same information for the flow distribution model. The computational complexity in each case is estimated using a fit (shown as dashed lines) of the type $Y=k X^\alpha$, where $k$ is an overhead constant, and $\alpha$ measures the computational complexity of the algorithm. The shaded region around each point represents its variance based on 20 samples.\label{fig4}}
\end{figure*}

\begin{figure}[t]
    \centering
    \includegraphics[width=0.5\textwidth]{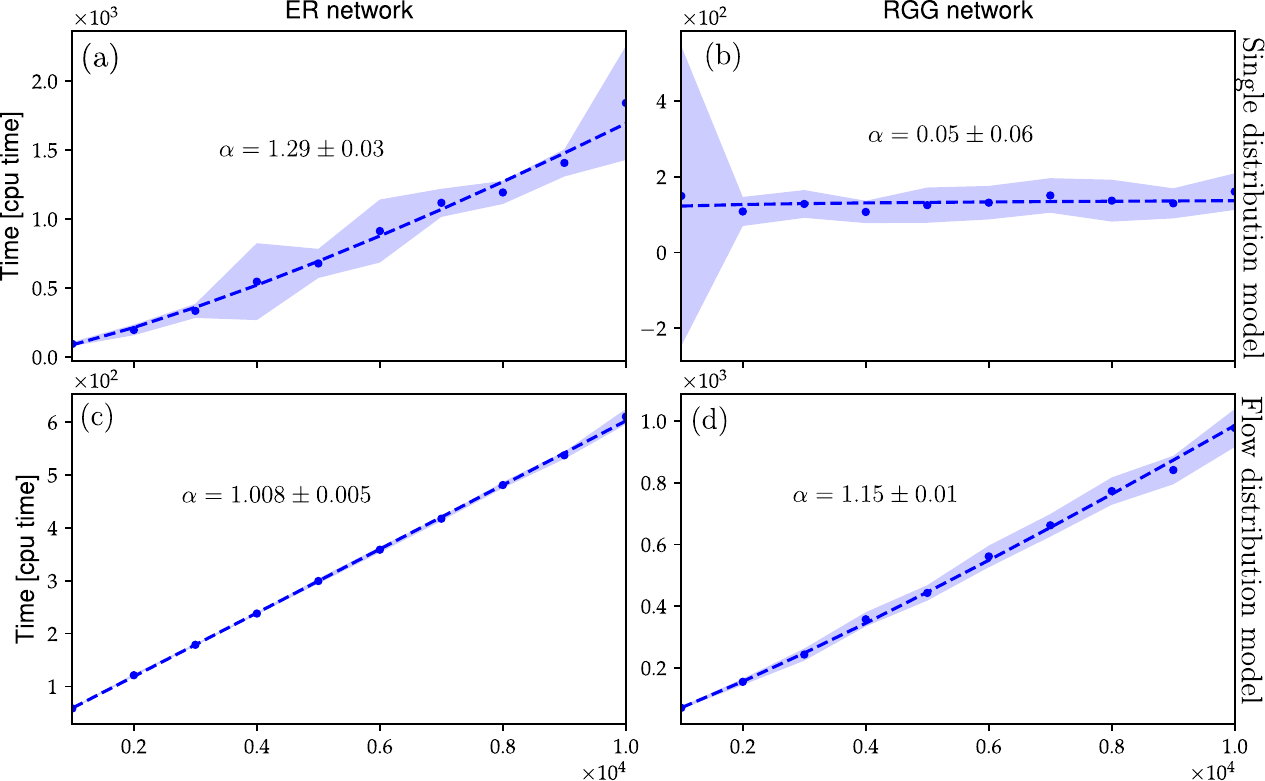}
    \caption{{\bf Multipartite routing runtime analysis}. Simulation of three-partite entanglement distribution for photonic quantum networks in an \er network (ER) and a random geometric graph (RGG) topology, an average degree $\langle k \rangle=10$ for the single, (a-b), and flow, (c-d), distribution models.  Each link follows the link model from \cite{Childress:2005} and his characterized by three parameters: $\epsilon$, the dark count $p_{\rm dark}$, and the minimum infidelity $\bar{F}_{max}$, randomly assigned from uniform distributions $U(0.3, 0.4)$, $U(0,{10}^{-3})$, and $U(0,{10}^{-3})$, respectively. . (a-b) show the total execution time of the algorithm as a function of the number of nodes for the same  networks. (b-d) show the same information for the flow distribution model. The computational complexity in each case is estimated using a fit (dashed lines) of the type $Y=k X^\alpha$, where $k$ is an overhead constant, and $\alpha$ measures the computational complexity of the algorithm. The shaded region around each point represents its variance based on 15 samples. 
    \label{fig5}}
\end{figure}

Numerical simulations were performed to find the best route for the distribution of bipartite Fig.~\ref{fig4} and multipartite entanglement distribution (Fig.~\ref{fig5}) for large complex networks using the link model described in \cite{Childress:2005}.  In such setups, the entanglement between qubits is generated by means of laser pulses. An increase in the duration of the laser pulses increases the probability of generating an ebit, but it also reduces its quality, as represented in Fig.~\ref{fig1}(a,b). For this reason, if the objective for a link is to generate a large amount of qubits, while its quality is a matter of less importance, one would use a longer laser pulse; in contrast, if the link requires just a few ebits but with high fidelity, then shorter pulses would be preferred. Let us look in more detail at the entanglement generation scheme presented by Childress~\cite{Childress:2005}, in which two NV centers are inside a photonic cavity, one at each end of a photonic fiber. Laser pulses are used to create entangled NV center excited state pairs with a fidelity.
\begin{align}
    F=&\frac{1}{2}\left(1+e^{-p_{\rm em}(1-\epsilon)}\right)-\frac{p_{\rm dark}}{p}-\bar{F}.
    \label{eq:F_vs_pem}
\end{align}
where $F$ is the fidelity of the generated ebit, and $p$ is the success probability. $p_{\rm em}$ is the emission probability that will increase with the duration of the applied laser pulse. $\epsilon$ is the collection efficiency, i.e., the probability of a photon being collected by the cavities. $p_{\rm dark}$ is the dark-count probability, referring to the probability of detecting a photon when none was emitted. $\bar{F}$ is a an upper bound for the infidelity and whose details are not directly relevant to our work; in practice, it limits the maximum fidelity that an entangled qubit pair can achieve using this setup. Finally, the success probability of generating an ebit is \cite{Childress:2005}

\begin{equation}
    p=\frac{1-e^{-p_{\rm em}\epsilon/2}}{2},
    \label{eq:p_vs_pem}
\end{equation} 

This trade-off between the fidelity and the success probability of generating an ebit gives rise to the aforementioned fidelity curves, which, for each link, characterize the trade-off between fidelity and capacity.Two types of network topologies were considered:
\begin{enumerate}[label=\roman*)]
    \item  an Erdős-Rényi (ER) network composed of $V$ nodes and $L$ links connecting two random nodes, which is a model that captures some properties of real-world networks, such as randomness and small-worldness;
    \item a random geometric graph, which models a network embedded in a geometric space, usually a d-dimensional box of volume 1, where $N$ nodes are randomly distributed. Two nodes are connected if the distance between them is smaller than a neighborhood radius r. This model captures a real limitation of today's technology for distributing entanglement directly over large distances.
\end{enumerate}
When using the proposed approach, the paths found always converge to the optimal solution in polynomial time, as shown in Fig.~\ref{fig4} and Fig.~\ref{fig5}. This is in line with multi-objective routing literature that, as mentioned before, indicates that while the problem itself is NP-hard, it often converges to the optimal solution in polynomial time \cite{bokler2017multiobjective}, making this approach feasible for quantum network routing optimization.

\section{Relation to previous works, Limitations, and Future directions}

With the surge of research on entanglement distribution and the wide range of models being considered, it is valuable to identify models that can be addressed using the proposed approach, which can be interpreted as a generalization of previous methods for single and multi-objective optimization. It evolves from discrete variables that seek the best path between a source and a set of nodes, to one based on continuous variables \cite{sarasantos2023, Bugalho2023, Pirandola2019, Bauml2020}.
%This approach offers the significant advantage of allowing us to use fidelity curves directly, thereby expanding the range of existing quantum network models in which one can employ routing based on multi-objective optimization.
However, it is important to note the limitations of the proposed algorithm. It cannot be used to find the combination of paths that maximizes the entanglement distribution rate from a source to a target \cite{chakraborty2020entanglement, Bauml2020, Pirandola2019} using multi-paths. Multi-path routing is often addressed using a linear programming formulation, which provides several advantages \cite{Pirandola2019, chakraborty2020entanglement, Bauml2020}. It allows multi-path routing, and can also be applied to scenarios involving multiple sources and multiple targets.
Unfortunately, to the best of our knowledge, our problem cannot be formulated as a linear programming one. This represents the biggest drawback of using linear programming for entanglement routing: the amount of detail one can add becomes restricted by the need to formulate the problem as a linear optimization. Nevertheless, the proposed approach allows for the addition of as much detail as needed, provided that monotonic and isotonic routing metrics can still be defined. An interesting way to merge these two directions would be to reformulate this work as a non-linear programming problem.
In its present form, the proposed technique cannot be used for entanglement distribution flow models considering non-deterministic swapping protocols because in that case the distribution rate will depend on the swapping order \cite{ghaderibaneh2022efficient}, making the problem considerably harder. Nevertheless, the proposed approach remains an extremely versatile tool that can be used in combination with relatively complex quantum network models. %In terms of future directions, it would be interesting not only to consider  non-deterministic swapping protocols but also to include purification.
%Purification combines multiple ebits pairs to generate a lower number of qubits with higher fidelity. Complex repeater protocols use rounds of  purification and swapping in order to distribute entanglement over large distances. To do this one needs to consider the different order in which the purification and swapping are applied, making the problem much harder to track, and likely it will  require a more complex routing algebra.  We leave the incorporation of non-deterministic swapping  and complex quantum repeater protocols for future works. 

\section{Conclusions}
This paper focused on multipartite entanglement distribution for a quantum network connected through links that exhibit a trade-off between entanglement generation rate and fidelity. This is the case with hash-based quantum purification protocols~\cite{Roque2023} and with photonic models~\cite{Childress:2005}. Two entanglement distribution models were considered: one where only one ebit is sent at each time epoch, and a second so-called flow model, where a large number of ebits are distributed simultaneously. The paper proposed using fidelity curves as a routing metric in both scenarios in combination with a multi-objective optimization algorithm, which finds the best path (or best star) connecting two (or three) nodes in close to linear time. The proposed method can be readily adapted to address routing challenges in various quantum network models, including those incorporating purification protocols between adjacent nodes. Nevertheless, how to deal with multi-path routing with non-deterministic swapping is still an open problem. In conclusion, this work paves the way for entanglement distribution in networks with complex link models, incorporating highly efficient purification protocols, and enabling optimization of quantum routing in more realistic and sophisticated network scenarios.

\section*{Acknowledgements}

This work was supported in part by the European Union’s Horizon 2020 Research and Innovation Program through the Project Quantum Internet Alliance (QIA) under Grant 820445, and in part by FCT - Fundação para a Ciência e Tecnologia, I.P. by project references QuNetMed 2022.05558.PTDC, with DOI identifier \url{https://doi.org/10.54499/2022.05558.PTDC}, and UIDB/50008/2020, with DOI identifier \url{https://doi.org/10.54499/UIDB/50008/2020}.
Raul Monteiro was supported by Fundação Calouste Gulbenkian through the Program New Talents in Quantum Technologies. Luís Bugalho acknowledges the support of FCT through scholarship BD/05268/2021.

\bibliographystyle{IEEEtran}
\bibliography{sample.bib}

\vskip 0pt plus -1fil
%\vskip 0pt
\begin{IEEEbiographynophoto}{Bruno C. Coutinho} obtained his PhD in Physics from Northeastern University, USA, in 2016, and both his BSc. and MSc. in Physics, from the University of Aveiro, Portugal, in 2009 and 2011, respectively. Since 2017 he is with the Physics of Information and Quantum Technologies Group, at Instituto de Telecomunicações, initially as a postdoc and later as a Research Fellow.
\end{IEEEbiographynophoto}

\vskip 0pt plus -1fil
\begin{IEEEbiographynophoto}{Raul Monteiro} obtained his MSc in Physics Engineering from Instituto Superior Técnico (IST), University of Lisbon, Portugal. He is currently a PhD student in Electrical and Computer Engineering, also at IST. He was a Calouste Gulbenkian Scholar in 2019/2020.
\end{IEEEbiographynophoto}

\vskip 0pt plus -1fil

\begin{IEEEbiographynophoto}{Luís Bugalho} obtained his MSc in Physics Engineering from Instituto Superior Técnico (IST), University of Lisbon, Portugal. He is currently a PhD student in Physics, also at IST. Presently, he is also a visiting research student at LIP6, Sorbonne Université, CNRS, Paris, France.
\end{IEEEbiographynophoto}

\vskip 0pt plus -1fil
\begin{IEEEbiographynophoto}{Francisco A. Monteiro} (M'07) is Associate Professor in the Dep. of Information Science and Technology at Iscte - University Institute of Lisbon, and a researcher at Instituto de Telecomunicações, Lisbon, Portugal. He holds a PhD from the University of Cambridge, UK, and the Licenciatura and MSc degrees in Electrical and Computer Engineering from Instituto Superior Técnico (IST), University of Lisbon, where he also became a Teaching Assistant. He held visiting research positions at the Universities of Toronto (Canada), Lancaster (UK), Oulu (Finland), and Pompeu Fabra (Barcelona, Spain). He has won two best paper prizes awards at IEEE conferences (2004 and 2007), a Young Engineer Prize (3rd place) from the Portuguese Engineers Institution (Ordem dos Engenheiros) in 2002, and for two years in a row was a recipient of Exemplary Reviewer Awards from the IEEE Wireless Communications Letters (in 2014 and in 2015). He co-edited the book ``MIMO Processing for 4G and Beyond: Fundamentals and Evolution'', published by CRC Press in 2014. In 2016 he was the Lead Guest Editor of a special issue on Network Coding of the EURASIP Journal on Advances in Signal Processing. He was a general chair of ISWCS 2018 - The 15th International Symposium on Wireless Communication Systems, an IEEE major conference in wireless communications.
\end{IEEEbiographynophoto}
\appendix

\subsection{Optimally of sequential concatenation of fidelity curves in the single ebit distribution model}
\label{optimality_0}

We want to prove that the sequential concatenation of fidelity curves in the single ebit distribution model always leads to the optimal solution. In a network, nodes are typically labeled by numbers, and pairs of nodes are used to identify edges, but here, to simplify the notation, we will consider a path $P$ with edges labeled numerically as $0,1,2,3,...L-1$, where $L$ is the length of the path. The optimal solution is described by:

\begin{align}
    \gamma_{P}^{\rm sg}(c^{\rm sg})&=\max_{\{c_{i}\}} \prod_{i =0}^{L-1}
    \gamma_{i}(c_{i}), \label{Eq:iso1}
\end{align} 
with the constraint 
\begin{align}
    c^{\rm sg}=\prod_{i =0}^{L-1} c_{i}.
\end{align} 
This constraint can be easily incorporated in Eq.~(\ref{Eq:iso1}), by rearranging the constraint as $c_0=c^{\rm sg}/\prod_{i =1}^{L-1} c_{i}$, and we obtain,

\begin{align}
    \gamma_{P}^{\rm sg}(c^{\rm sg})&=\max_{\{c_i\}} \left[ \gamma_{0}\left(\frac{c^{\rm sg}}{\prod_{i =1}^{L-1}c_{i}}\right) \prod_{i =1}^{L-1} \gamma_{i}(c_{i})\right]\label{Eq:iso2},
\end{align} 
where we added the constraint to edge $0$, but one can add the constraint to any other edge. Now that we know how to write the optimal solution for the fidelity curve in {Eq.~(\ref{Eq:iso1})}, let us consider the concatenation of a path $0$ and a path $1$, which can be written as,
\begin{align}
   \gamma^{\rm sg}(c_{01}^{\rm esg})  &=\max_{c_{0},c_{1}} \left[ \gamma_{0}(c_{0}) \gamma_{1}(c_1)   \right] \label{Eq:iso3}
\end{align} 
with the constraint
 \begin{align}
     &c_{01}^{\rm sg}:=c_0c_1. 
\end{align} 

Using the same approach as before, we can include this constraint into Eq.~(\ref{Eq:iso3}), and obtain

\begin{align}
   \gamma^{\rm sg}(c_{01}^{\rm sg})  &=\max_{c_{1}} \left[ \gamma_{0}\left(c_{01}^{\rm sg}/c_1\right) \gamma_{1}(c_1)   \right]. \label{Eq:iso2}
\end{align} 
Using the new notation for paths defined above, the concatenation of path $01$ with edge $2$ is given by
\begin{align}
   \gamma^{\rm sg}(c_{012}^{\rm sg})  =&\max_{c_1} \left[ \gamma_{01}\left(\frac{c_{012}^{\rm sg}}{c_2}\right) \gamma_{2}(c_2)   \right]= \\
                                     &\max_{c_{1}} \left[ \gamma_{0}\left(\frac{c_{012}^{\rm sg}}{c_1 c_2}\right) \gamma_{1}(c_1)\gamma_{2}(c_2)\right].
   \label{Eq:iso2}
\end{align} 
Repeating this process until edge $L-1$, one obtains Eq.~(\ref{Eq:iso3}), proving the sequential concatenation of paths in the single ebit distribution model leads to the optimal fidelity curve.

\subsection{Optimal link capacity in the flow distribution model}
\label{optimality_1}
In this section, we determine the optimal capacity of each link in a quantum repeater chain for our model. In general, we can write a system of two equations to describe the fidelity in the flow distribution model, as
\begin{align}
    &\gamma_{P}(c_{P})=\max_{\{c_e\}}\prod_{e \in P} \gamma_{e}(c_{e}), \label{eq:appendex_0} \\
    &c_{P}=\min_{e \in P} c_{e}   \label{eq:appendex_1}
\end{align} \par 
\noindent where $P$ is a path connecting nodes $i$ and $j$.
One wants to prove that this fidelity curve is a monotonically decreasing function, and the optimal solution to this set of equations, i.e., a solution that maximizes the Werner parameter $\gamma_{P}$ for a given rate $r_{P}$, is for all links to operate at a capacity equal to the entanglement distribution rate $c_{e}=c_{P}$, for all values of $e$. Let us call this the uniform solution. We will now prove that the uniform solution is optimal using proof-by-contradiction. Let us consider a solution of Eqs.~(\ref{eq:appendex_0})-(\ref{eq:appendex_1}) $\{c'_e\}$, with the same entanglement distribution rate as the uniform solution $c_{e}=c_{P}$, but a larger Werner parameter, i.e.,
\begin{align}
    \prod_{e \in P} \gamma_{e}(c'_{e})> \prod_{e \in P} \gamma_{e}(c_{P})\\\implies
    \prod_{e \in P} \frac{\gamma_{e}(c'_{e})}{\gamma_{e}(c_{P})}> 1
    \label{contradic:appendex_1}
\end{align} 
Since the entanglement distribution of our optimal solution is $c_{P}$, we know that at least one link operates at capacity $c_{P}$, and all others operate at a capacity equal \textcolor{blue}{to} or larger, $c'_{e}\leq c_{P}$. Assuming that fidelity curves are monotonically decreasing functions (as we do in the work), the $\gamma$-value of a link operating at a capacity $c_{P}$ is larger or equal to the $\gamma$-value of that same link operating at a capacity $c'_{e}$. One can conclude that $\gamma_{e}(c'_{e})\leq \gamma_{e}(c_{P}) \implies \gamma_{e}(c'_{e})/\gamma_{e}(c_{P})\leq 1$ for all $e$ in $P$, contradicting Eq.~(\ref{contradic:appendex_1}).

\end{document}